\documentclass{emulateapj} \usepackage{psfig} \usepackage{apjfonts}
\usepackage{graphicx}

\newcommand\beq{\begin{equation}}
\newcommand\eeq{\end{equation}}
\newcommand\lsim{\mathrel{\rlap{\lower4pt\hbox{\hskip1pt$\sim$}}
        \raise1pt\hbox{$<$}}}
\newcommand\gsim{\mathrel{\rlap{\lower4pt\hbox{\hskip1pt$\sim$}}
        \raise1pt\hbox{$>$}}}
\newcommand\axp{1E~2259+586}
\newcommand\psr{PSR~J1814-1744}

\begin{document}

\title{Magnetars vs. high magnetic field pulsars: a theoretical 
interpretation of the apparent dichotomy}

\author{Jose A. Pons\altaffilmark{1} and Rosalba Perna\altaffilmark{2}} 
\altaffiltext{1}{Department de Fisica Aplicada, Universitat d'Alacant,
Ap. Correus 99, 03080, Alacant, Spain}
\altaffiltext{2}{JILA and Department of Astrophysical and Planetary Science,
University of Colorado at Boulder, 440 UCB, Boulder, CO, 80304}

\begin{abstract}

Highly magnetized neutron stars (NSs) are characterized by a
bewildering range of astrophysical manifestations.  Here, building on
our simulations of the evolution of magnetic stresses in the NS crust
and its ensuing fractures (Perna \& Pons 2011), we explore in detail,
for the middle-age and old NSs, the dependence of starquake frequency
and energetics on the relative strength of the poloidal ($B_{\rm p}$)
and toroidal ($B_{\rm tor}$) components.  We find that, for $B_{\rm
  p}\ga 10^{14}$~G, since a strong crustal toroidal field $B_{\rm
  tor}\sim B_{\rm p}$ is quickly formed on a Hall timescale, the
initial toroidal field needs to be $B_{\rm tor} \gg B_{\rm p}$ to have
a clear influence on the outbursting behaviour. For initial fields
$B_{\rm p}\la 10^{14}$~G, it is very unlikely that a middle-age ($t
\sim10^5$ years) NS shows any bursting activity.  This study allows us
to solve the apparent puzzle of how NSs with similar dipolar magnetic
fields can behave in a remarkably different way: an outbursting
'magnetar' with a high X-ray luminosity, or a quiet, low-luminosity,
'high-$B$'' radio pulsar.  As an example, we consider the specific
cases of the magnetar \axp\, and the radio pulsar \psr, which at
present have a similar dipolar field $\sim 6\times 10^{13}$~G. We
determine for each object an initial magnetic field configuration that
reproduces the observed timing parameters at their current age.  The
same two configurations also account for the differences in quiescent
X-ray luminosity and for the 'magnetar/outbursting' behaviour of
\axp\, but not of \psr.  We further use the theoretically predicted
surface temperature distribution to compute the light-curve for these
objects. In the case of \axp, for which data are available, our
predicted temperature distribution gives rise to a pulse profile whose
double-peaked nature and modulation level is consistent with the
observations.

\end{abstract}


\section{Introduction}

Neutron stars, end points of the evolution of massive stars, are
characterized by a bewildering variety of astrophysical
manifestations.  Among these, particularly intriguing is a class of
slow rotators ($P\sim 2-11$~sec) with an especially high quiescent
X-ray luminosity ($L_x\sim 10^{33}-10^{35}$~erg), almost always exceeding
their rotational energy loss $\dot{E}$, and displaying occasional
energetic outbursts and giant flares. The latter can release energies
as large as $\sim 10^{45}$~erg.

The source of energy responsible for powering the outbursts and flares, and
enhancing the quiescent emission, is believed to be of magnetic origin
(Thompson \& Duncan 1995, 2001). The magnetic field evolves 
in the NS interior under the effects of Ohmic dissipation, Hall drift and
Ambipolar diffusion, although this latter has recently been suggested to 
play a lesser role in the presence of superfluid matter  \citep{Glam11}.
In this case, the evolution of the  magnetic field causing
the magnetar activity would likely take place in the stellar crust.
However, if the observational evidence from the cooling of the supernova remnant
in Cassiopeia A \citep{casa1,casa2} is confirmed, it would imply that the object is
undergoing a core superfluid transition at $5\times 10^8$ K. In the magnetar context,
higher core temperatures are expected and the transition to core superfluidity can be significantly delayed,
as originally suggested by \cite{arras04}. Therefore, the role of ambipolar diffusion in the core
remains an open question. In this work we focus on the role of Ohmic diffusion and Hall drift
in the crust.

Occasionally the magnetic stress exceeds locally the tensile
strength of the crust, causing it to fracture; the ensuing sudden 
release of elastic and magnetic energy powers the outbursts.  
This picture, in which the magnetic field strength 'regulates' the
properties of an object, i.e. its appearance as an outbursting  magnetar, or a  
'normal', quiet radio pulsar, left many unanswered questions and unsolved
puzzles. Why some objects are more active than others? What 
determines the frequency and energetics of events? Why some objects
have been observed to emit very energetic  $\gamma$-ray flares (the so-called ``Soft 
Gamma-ray Repeater'', (SGRs)), while others only display less
energetic X-ray bursts (the so-called ``Anomalous X-ray Pulsars'' (AXPs))?
Is there a real difference between these two classes ?
 And, if the magnetic field is the main driving force, how is it
possible for a ``low $B$''-field pulsar to undergo an outburst (Rea et al. 2010)?
Altogether, how can two objects, characterized by an almost identical 
magnetic field (as inferred through measurements of the pulsar period $P$ and
its derivative $\dot{P}$) manifest themselves as apparently different objects
with respect to those properties which should be driven by the $B$-field strength
(outbursting behaviour, quiescent X-ray luminosity)?
This apparent dichotomy is exemplified by the magnetar \axp, with an inferred dipolar
field of $5.9\times 10^{13}$~G, and the 'high-$B$' field radio pulsar \psr,
with an inferred $B_{\rm dip}=5.5\times 10^{13}$~G. The former is classified as an
AXP; it was observed to have an active period of outbursts in 2002 (Woods et al. 2004),
and has an enhanced X-ray luminosity in quiescence, $L_x\sim 10^{34}$~erg, which
exceeds its rotational energy loss. On the other hand, \psr\, appears to be a ``normal'', quiet radio
pulsar (Camilo et al. 2000), with no discernible X-ray emission in the X-ray band.

\begin{figure*}[ht]
\includegraphics[angle=0]{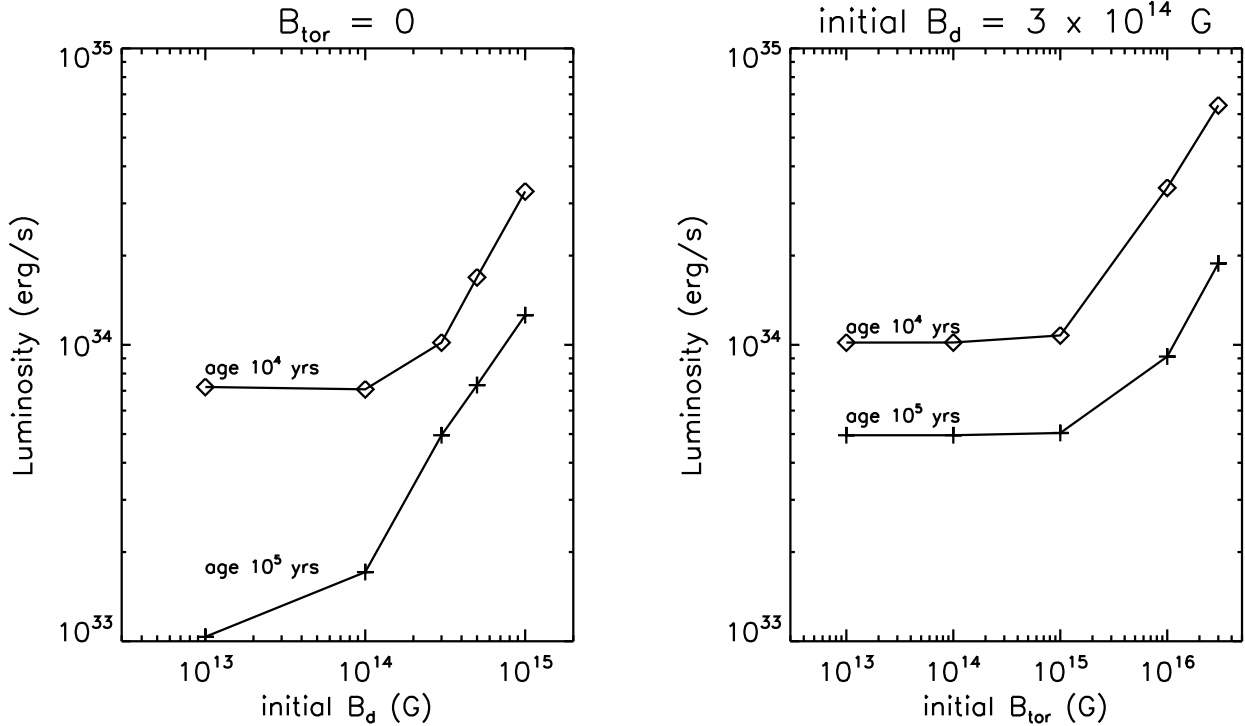}
\caption{Luminosity at two different ages as a function of the initial magnetic field.
Left panel: dependence of the luminosity on the initial poloidal field (at pole) for
models with $B_{\rm tor}=0$. Right panel: same for models with $B_{\rm p}=3\times 10^{14}$~G
and different strength of the initial toroidal field. 
\label{fig1}}
\end{figure*}

In order to understand from first principles what is the physical origin of the
variety of observed phenomenology, in Perna \& Pons (2011; PP2011 in
the following), we performed the first long-term 2D simulations that
followed the evolution of magnetic stresses in the NS crust; these,
combined with calculations of the breaking stress of the NS, allowed
us to theoretically predict frequency, energetics, and location on the
NS surface of the outbursts (associated with crust fractures). Our
results allowed us to establish that there is no fundamental
difference among apparently different manifestations of the objects
(such as 'AXPs' and 'SGRs'). We also found that both the dipolar component and
the 'hidden' toroidal component play an equally important role in
determining the  frequency and energetics of the starquakes. Furthermore,
for a given initial $B$-field configuration and strength, younger objects
yield more frequent and more energetic outbursts.

In this follow up paper, we continue our work in several
respects. Firstly, we extend our study of the outburst frequency and energetics 
to focus on the influence of the relative strengths of the poloidal and
toroidal field components. This study allows us to uncover some
fundamental clues as to what creates the varied phenomenology of the highly magnetized
neutron star family. Second, for the middle-aged and older objects, we
consider the mutual influence of the magnetic and thermal evolution and predict, for a
combination of $B_{\rm p}$ and $B_{\rm tor}$ initial strengths, the X-ray
luminosity, temperature, timing properties and magnetic field at current age. 
Last, as an example, we consider two specific objects, \axp\, and \psr, which, as described above, have
almost identical spin parameters and hence inferred dipolar $B$ field,
but very different manifestations.  For each of them, we
show how we can find an initial $B$-field configuration that
reproduces their observed timing parameters at their current timing
age, as well as account for the differences in quiescent X-ray
luminosity, and for the 'magnetar/outbursting' behaviour of \axp\, but
not \psr. We also compute the surface temperature distributions associated with
the magnetic configurations that reproduce those properties and use
them to compute the expected X-ray light-curves
in quiescence. For the case of \axp, for which data are
available, we show that the qualitative features that we predict are a
good match to the data.

The paper is organized as follows.
The magneto-thermal evolution is described in
\S2, together with predictions for the intensity of the X-ray
luminosity with age as a function of $B_{\rm p}$ and $B_{\rm tor}$. 
In \S3, we study the dependence of the starquake frequency and energetics on the relative
strengths of $B_{\rm p}$ and $B_{\rm tor}$.  The specific
examples of magneto-thermal evolution for \axp\, and\psr\, are
presented in \S4. We summarize and discuss our work in \S5.

\section{Magneto-thermal evolution of highly magnetized neutron stars}

The first attempt to study the long term cooling of magnetized NSs,
including all realistic microphysics and taking into account magnetic
field decay was made by \cite{aguil08}. They assumed a
simple, analytical law for the time variation of $B_{\rm p}$ which
incorporates the main features of the Ohmic and Hall terms in the
induction equation.  Numerical simulations of the magnetic field
evolution in a realistic neutron star crust, including the nonlinear
effects of the Hall term were presented in \cite{PG07}, but in this
case assuming an isothermal crust.  The fully coupled magneto-thermal
evolution of a NS was later studied by \cite{pons09}. However, owing
to numerical difficulties in treating the Hall term, their models
could include only Ohmic diffusion. A fully coupled evolutionary
simulation has not been possible up to now, but for initial values of
$B_{\rm p}\lesssim 10^{15}$~G, the effect of the Hall drift is expected to
introduce at most quantitative changes (a somewhat faster dissipation)
with respect to the purely Ohmic picture described in \cite{pons09},
which can be considered a good approximation to reality.  With this
code, we performed a number of simulations of our baseline model (a
$1.4M_\odot$ NS with a radius of 11.6~km, and moment of inertia $I=1.4
\times 10^{45}$~g cm$^2$) varying the initial values of the dipolar
and internal toroidal fields.  We refer to \cite{pons09} and
\cite{aguil08} for all technical details about the code and the
microphysical input. The minimal cooling scenario \citep{page04},
without exotic phases nor fast neutrino cooling processes, was
assumed. We also include enhanced neutrino emission from the breaking
and formation of neutron Cooper pairs in the core, consistently with
the observational evidence from the cooling of the supernova remnant
in Cassiopeia A \citep{casa1,casa2}.

\begin{figure}
\includegraphics[width=3.2in,angle=0]{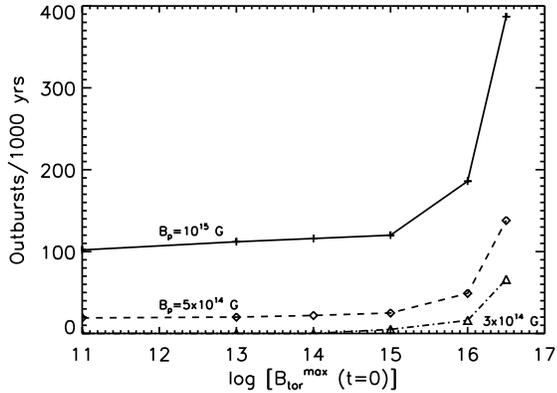}
\caption{Average number of outbursts per 1000 years as a function of the initial
toroidal field strength, for a $10^5$ years old neutron star. We show results for three different initial poloidal fields:
$B_{p}(t=0)=10^{15}$~G (solid lines), $B_{p}(t=0)=5\times10^{14}$~G (dashes), and
 $B_{p}(t=0)=3\times10^{14}$~G (dash-dotted lines).
\label{fig2}}
\end{figure}


As noted in previous studies, the average luminosity of a NS born with
$B_{\rm p} < 5 \times 10^{13}$ G.  is hardly affected, compared to the
non-magnetized case.  On the other hand, those NSs born as magnetars
are subject to significant heating by the dissipation of currents in
the crust. At fixed poloidal field, the luminosity of models with
strong internal toroidal components is systematically higher than that
of models without toroidal fields, due to the additional energy
reservoir stored in the toroidal field, which is gradually released as
the field dissipates.  In Fig.~\ref{fig1} we show the dependence of
the luminosity at a fixed age as a function of the initial magnetic
field.  We can see that, for low initial dipolar fields, the
luminosity barely changes with increasing dipolar field but, once we
enter in the magnetar range, $B_{\rm p} > 10^{14}$ G, a further increase in
$B_{\rm p}$ results in a luminosity increase of more than one order of
magnitude. At fixed poloidal field, the existence of an internal
toroidal field has no evident effect if $B_{\rm tor} \lesssim B_{\rm p}$,
but the luminosity is again largely increased when the internal
toroidal field exceeds $B_{\rm p}$.
This can be understood with the following back-of-the-envelope estimate. The rate at which
magnetic energy $e_m$ is dissipated can be approximated by
\beq
\frac{d e_m}{dt} \approx \frac{J^2 V}{\sigma}
\eeq
where $J=\frac{c}{4 \pi} {\nabla \times B}$ is the current density, $V$ is the volume where currents are dissipated (the crust volume)
and $\sigma$ is the electrical conductivity. Approximating $\nabla \times B \approx B/L$, where $L$ is a typical scale-length,
and for typical neutron star conditions we have, for $L=1$~km, that 
\beq
\frac{d e_m}{dt} \approx 10^{33} \frac{B_{15}^2}{\sigma_{24}} {\rm erg/s}.
\eeq
\noindent
Here $B_{15}$ is the magnetic field in units of $10^{15}$ G and $\sigma_{24}$ is the electrical
conductivity in units of $10^{24}$ s$^{-1}$.
In order for the internal field to have a significant effect on the luminosity, the dissipation rate of magnetic
energy has to be larger than the luminosity in the non-magnetic case. This explains the right panel of Fig.~\ref{fig1},
where the influence of the toroidal field only becomes visible after a critical value of $10^{15}$ G and then
increases rapidly with the field strength.

We must remind that the quoted values of $B_{\rm p}$ and $B_{\rm tor}$
correspond to the initial values.  In the models with stronger fields,
as a result of the average higher temperatures, the crustal electrical
resistivity is also enhanced and magnetic diffusion is more efficient
during the first $10^5-10^6$ years of a NS's life. As a consequence,
all NSs born with fields $\ga 5 \times 10^{14}$~G are subject to
faster dissipation.  Typically, the dipolar field decreases by a
factor of 2-3 in the first $10^5$ years of a NS life, and it
asymptotically reaches a similar field strength ($\approx 2-3 \times
10^{13}$~G) at very late times, irrespectively of the initial
strength. The toroidal field at $10^5$ years is reduced (on average)
by one order of magnitude with respect to the initial value. 
Note also that, even if the initial model has no toroidal field, for sufficiently large
(poloidal fields $\ga 5 \times 10^{14}$~G) the Hall drift will create a toroidal field of similar strength.

\begin{figure}
\includegraphics[width=3.2in,angle=0]{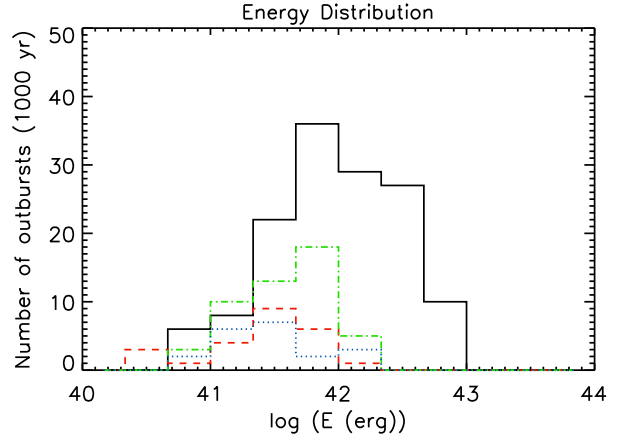}
\caption{Outburst energy distribution of objects at age $10^5$ yr born with
$B_{\rm p}=5\times 10^{14}$~G and four different initial toroidal fields: 
$B_{\rm tor}=0$ (blue dots), $10^{15}$~G (red dashes), $10^{16}$~G (green dash-dotted line), and
$3\times 10^{16}$~G (black solid line). The total number of outbursts corresponds to a period
of 1000 yr.
\label{fig3}}
\end{figure}

\section{Outburst frequencies as a function of the strength of the dipolar and toroidal
$B$-field components}

In PP2011, we used the numerical code by
\cite{PG07} to follow in axial symmetry the evolution of
the magnetic field in the crust of the neutron star. The magnetic
stress $M_{ij}(r,\theta,t)=B_i(r,\theta,t)B_j(r,\theta,t)/4\pi$ is
computed at each time as a function of the latitude $\theta$ on the
surface of the star, and the radial coordinate $r$ in the NS crust.  
As the magnetic field evolves, the crust moves through a
series of equilibrium states, during which elastic stresses
balance variations in magnetic stresses. If, during the evolution, any component
of the magnetic stress departs from its equilibrium value by an
amount comparable (or exceeding) the breaking stress of the crust,
the crust will break, releasing the accumulated elastic energy (for
technical details of the calculations, see PP2011). The simulations allowed PP2011 to predict the
frequency and energetics of the starquakes, for a certain initial
$B$-field configuration characterized by a poloidal and a toroidal
component. The case of $B_{\rm p}=8\times 10^{14}$~G and $B_{\rm tor}=2\times
10^{15}$~G was considered in detail in that work.  It was shown how
both the frequency and the energetic of the outbursts decrease with
the NS age. Some initial exploration of the role of
different initial field strengths further showed that the toroidal
component does play a very important role in the outburst properties.

The former study was limited by the numerical restrictions of the code that
do not allow to perform a systematic study of the starquake frequency 
and energetics of objects with $\gtrsim 10^5$~yr with the non-linear terms (too long runs).  
In order to overcome the numerical difficulties  we have adopted a hybrid approach.
Since, for these objects, the Hall time is much shorter than their age, we expect that the
initial configuration relaxes to a nearly {\it Hall-neutral} geometry on a few Hall timescales
($t_{\rm Hall}=4\pi n_e e L^2/(c B) \sim 6000 B_{14}^{-1}$~yr), after which
the evolution becomes dominated by Ohmic diffusion \citep{PG07}. Thus, as a first approximation,
we follow the NS evolution up to $10^5$ yr with the code of \cite{pons09}, and then use the temperature
and magnetic field profiles as initial input for the short-term, fully non-linear 
evolution including the Hall term, using the numerical code by
\cite{PG07}. The caveat of this approach is that it does not account for possible fast, Hall-induced, initial changes
of the magnetic field geometry and strength that may occur for magnetar conditions. This can result in a
dissipation of the initial field in, at most, a factor of 2, as estimated in \cite{PG07}.
It is also important to note that, whatever the initial geometry of the toroidal field was, it is rapidly (Hall timescale)
driven to a similar shape that corresponds to a sort of quasi-steady state. Therefore, our results for
middle aged NSs ($\gtrsim 10^5$~yr) are expected to be rather insensitive to the particular form of the
toroidal field, being more relevant the role of global quantities (total energy or helicity) that are explored
by varying the normalization of the toroidal component.

Our main results concerning the outburst frequency and energetics are
summarized in Fig.~\ref{fig2} and Fig.~\\bfref{fig3}. In Fig.~\ref{fig2}
we show the dependence of the average number of expected outbursts in
a period of 1000 yr on the strength of the initial magnetic field. We
have taken the parameter $\epsilon =0.9$ (see PP2011).  We can see
that the average outburst rate is sensitive to both $B_{\rm p}$ and $B_{\rm
  tor}$ in a similar way as the luminosity. Below a critical poloidal
field $B_{\rm p}<3\times 10^{14}$~G the burst rate is vanishingly small ($<
10^{-3} $ yr$^{-1}$), except for very large toroidal fields $B_{\rm
  tor}>10^{16}$~G, but even in that extreme case the outburst
frequency is smaller than one per century.  For initial $B_{\rm p}>3\times
10^{14}$, the outburst rate grows exponentially with the value of
$B_{\rm p}$.  At a fixed $B_{\rm p}$, the dependence on the toroidal field is very
weak until $B_{\rm tor} \sim B_{\rm p}$, but the outburst rate again
increases rapidly with increasing values of $B_{\rm tor} > B_{\rm p}$.  It
should be noted that these numbers must be taken with caution, simply
as an order of magnitude estimate. The large uncertainties about the
internal geometry of the magnetic field and details of the crustal
fracture mechanisms do not allow a precise calculation. Nevertheless,
these estimates may serve as an indication of the average outburst
frequency for objects with similar conditions.  The energy
distribution of the events during a period spanning 1000 years is
shown in Fig. \ref{fig3}, for the model with $B_{\rm p}= 5\times 10^{14}$~G
and four different toroidal fields.  Except for the model with the
largest $B_{\rm tor}$, there is no significant variation in the
outburst average event energy, which is predominantly in the range
$10^{41}-10^{42}$ erg. Only for the extreme case of $B_{\rm tor}= 3
\times 10^{16}$~G, in addition to the large increase in the number of events,
the average outburst energy is also substantially larger, up to $10^{43}$
erg, reflecting the larger energy reservoir available in the toroidal field.

\section{\axp\, and \psr: similar $B_{\rm p}$ but different astrophysical manifestation}

\begin{figure*}[t]
\begin{center}
\includegraphics[width=6.4in,angle=0]{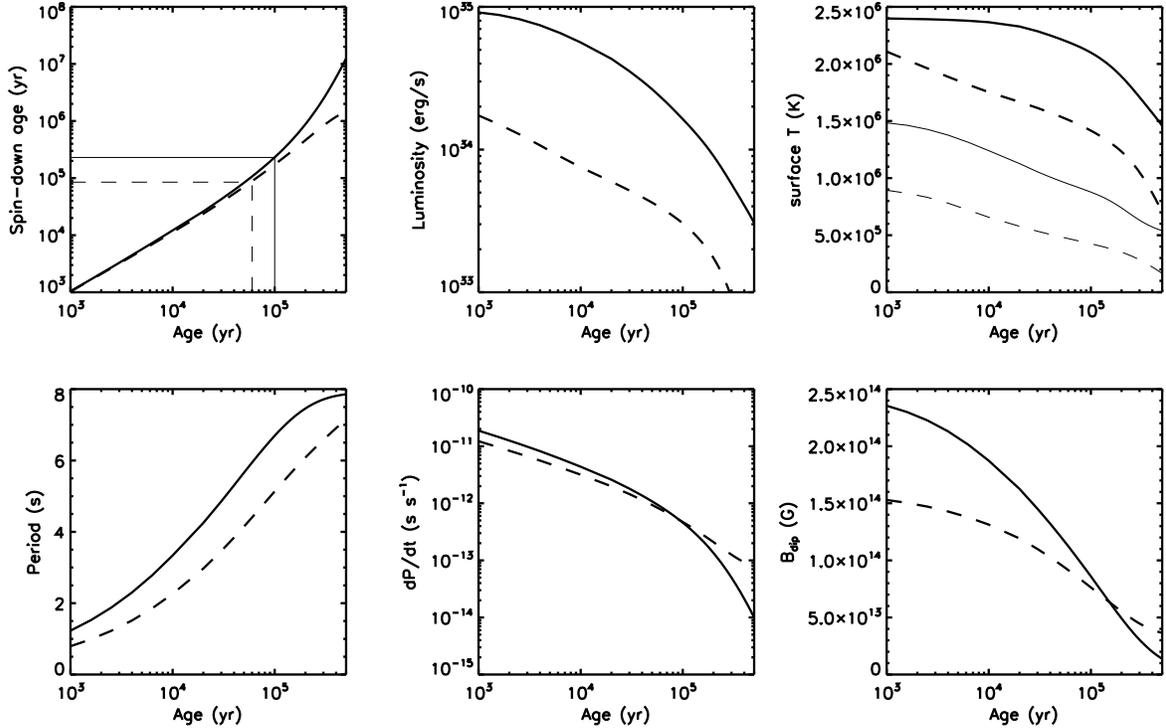}
\caption{Evolution of timing, thermal and magnetic variables for some
  choices of initial magnetic field configuration that reproduce the
  observed properties of \axp ~(solid line) and \psr ~(dashed line).
  The spin-down ages of 230~kyr and 85~kyr are reached at 
  real ages of 100~kyr and 60~kyr, respectively ({\em top left} panel).  
  In the upper {\em top right} panel the evolution of the redshifted temperature at the pole (thick) and
  the equator (thin) is shown for both models.
  \label{fig4}}
\end{center}
\end{figure*}
\noindent

The comprehensive study performed in the previous two sections allows
us to obtain a physical explanation to an apparent puzzle in neutron
star theory. How can two objects have a very similar dipolar field
but display a markedly different behaviour, if the magnetic field
provides the main driving force? 

Our magnetothermal simulations provide an answer to this question: for
a similar inferred value of the dipolar component of the magnetic
field, the strength of the internal toroidal field can have a dramatic
influence on the behaviour of the neutron star, and in particular on
its manifestation as an X-ray bright, outbursting 'magnetar', or an
X-ray dim, quiet 'high-$B$' radio pulsar.  In the previous section we
have shown that, if the {\em initial} dipolar component is low ($\la
\times 10^{14}$~G), the toroidal component, even if strong, is not
very effective in producing frequent outbursts.  On the other hand,
for dipolar field strengths on the order of a few $\times 10^{14}$~G,
a toroidal component of the same strength is produced during the
evolution, even if it was not there initially. Thus, the importance of
an existing initial $B_{\rm tor}$ is only evident when it exceeds the
poloidal component.  As our results have shown, over ages of about
$10^5$~yr, a dipolar $B$-field of a few $10^{14}$~G roughly halves.  It is
especially interesting to note that it is around dipolar magnetic
fields of $6-7 \times 10^{13}$~G (and ages of about $10^5$~yr) that
objects classified as 'magnetars' or 'high-$B$' radio pulsars are
found to coexist.  The most notable examples, as discussed in \S1, are
\axp, with an inferred $B_{\rm p}=5.9\times 10^{13}$~G and magnetar
characteristics, and \psr, with $B_{\rm p}=5.5\times 10^{13}$~G and
properties of a ``normal'' radio pulsar.  
We do not include other interesting objects in this study, such as the high magnetic
field radio pulsars PSR J1119-6127 or PSR J1846-0258, because they are very young
($\approx$ 1 kyr) and a fully coupled treatment of the Hall term with the thermal
evolution would be required. We have however also examined the high-$B$ radio pulsar
PSR~J1718-3718, and found that the properties of this object can be
well explained by the same model used for \psr, but in an earlier
stage. In the remaining of this section we specifically focus on the cases of \axp and
\psr and we leave a more extensive study of many different objects for the future. 

We evolved in time our baseline NS varying only the initial $B$-field configurations
(characterized by $B_{\rm p}$ and $B_{\rm tor}$) trying to reproduce the
timing, the magnetic, and the thermal properties of these two objects
at their measured timing age.  We assume an orthogonal rotator (angle
between the magnetic axis and rotation axis of $90^{\circ}$), and
initial spin period of 1 ms, short enough to have no influence on the
late-time timing properties. The two models that would represent the
two objects under discussion are presented in the following.

\bigskip
\noindent
{\em \underline{\axp.}}\\
\noindent

  This X-ray pulsar is characterized by a period $P=6.97$~s and period
  derivative $\dot{P}=4.8\times 10^{-13}$~s/s (Gavriil \& Kaspi 2002),
  implying a timing age of $\tau=230$~kyr. Its X-ray luminosity in the
  2-10 keV band of $3.4\times 10^{34}$~erg/s (for a distance of 4 kpc,
  Tian et al. 2010), much larger than its rotational energy, prompted
  its classification as an AXP.  The discovery of a major outburst in
  2002 (Kaspi et al. 2003) further tightened the connection between
  this AXP and the SGRs, and established its 'magnetar-like' nature,
  despite the low inferred dipolar field of $B_{\rm p}=5.9\times
  10^{13}$~G. The quiescent (pre-outburst) spectrum of this object is
  characterized by the superposition of a thermal and a power-law
  component.  The former, if fitted with a simple blackbody (BB),
  yields a BB temperature of about 0.4~keV and a BB radius of about
  5.3~km.  The pulse profile is double peaked (see Woods et al. 2004
  for a summary of the pre-outburst spectral properties). The
  modulation level (or pulsed fraction, PF) was measured in two XMM
  observations prior to the outburst (Woods et al. 2004); of interest here are the two
  softest bands (0.3-1.0 and 1.0-2.0 keV), dominated by the thermal
  component; the measured PF was found to be $0.169\pm 0.015$ in the
  0.3-1.0~keV band and
  $0.195\pm0.006$ in the 1-2~keV band in the first of the two
  observations, and $0.215\pm 0.006$ in the 0.3-1.0~keV band and
  $0.225\pm0.003$ in the 1-2~keV band in the second observation
  (quoted errors are $1\sigma$; note that there is some variability
between the measured values in the two observations). 

Fig. \ref{fig4} (solid lines) shows an example of such evolution. The
initial magnetic field has components $B_{\rm p} = 2.5\times 10^{14}$~G and
$B_{\rm tor}= 2.5\times 10^{16}$~G. In the upper left panel we see how
the timing age of the source increasingly departs from the real
age. At the measured spin-down age of 230~kyr, the real age of the
object is 100~kyr. At this age the predicted luminosity of the object
(in its thermal component, computed by integration of the temperature
profile over the whole surface of the star) is about $2\times
10^{34}$~erg/s, its period about 6.7~s, its spin-down rate about
$4.6\times 10^{-13}$~s/s, and its $B_{\rm p}$ about $6.7\times 10^{13}$~G.
Note that our realistic NS model described in Section 2 has a different radius and moment of
inertia from the canonical values (10 km, $10^{45}$~g cm$^2$) usually
assumed to infer $B_{\rm p}$.  Our goal is not to make a formal fit to the
properties of the object, but simply to show typical examples of
representative objects within a class.  Nevertheless, the quantities
in Fig.~1 are a very close match to the corresponding observed
properties of this object. For the thermal quantities (luminosity,
temperature), the comparison should however be made with care. The
measured luminosity is in fact subject to the uncertainty in distance,
and is also model-dependent.  Our simulations also predict the
temperature profile during the pulsar evolution.  We show the
evolution of the surface temperature at the pole and the equator in
the upper right panel of Fig.~4. A direct comparison between our
results and the observations can only be done at a qualitative level
at this stage. In fact, while our predictions specifically refer to the
thermal luminosity, the measured X-ray luminosity has both a thermal
and a non-thermal (powerlaw, PL) component (e.g. Woods et
al. 2004). The flux ratio PL/BB is a function of phase, varying
between 0.3 and 0.55 in the 0.5-7~keV energy band.  If the non-thermal
part is due to reprocessing of the thermal photons, then the
comparison should be made with the total luminosity, but if the
non-thermal photons are of different origin, then only the thermal
component should be considered.

In addition, the temperature inferred at infinity depends on both the
compactness ratio $M/R$ of the star, and the assumed model atmosphere,
which greatly affects the spectrum.  For the same NS compactness ratio
(and hence same gravitational redshift), fits with Hydrogen
atmospheres yield lower effective temperatures $T_{\rm atm}$. The
exact ratio $T_{\rm BB}/T_{\rm atm}$ depends on a number of factors,
such as atmosphere composition, magnetic field strength, temperature,
and, to a lesser extent, on column density and instrument type (see e.g. Suleimanov et al. 2011 
for recent work using non-magnetic model atmospheres with varying composition).  An
extensive exploration of $T_{\rm BB}/T_{\rm atm}$ for different values of the
above parameters was made by D. Lloyd with detailed magnetized
Hydrogen atmosphere models (Lloyd 2003).  For a $B$-field strength of
about $10^{13}$~G, a surface temperature of $1-2\times 10^5$~K, and a
column density $\sim 10^{21}$~cm$^{-2}$, he found the ratio $T_{\rm
  BB}/T_{\rm atm}$ to be $\approx 2$ (using the ACIS-1 spectra
simulator).  The observed spectrum of \axp\, has been fitted with
a BB+PL before the outburst (Woods et al. 2004), and with both a BB+PL
and a 2BB after the outburst, as it approaches quiescence (Zhou et
al. 2008). In the observation closest to quiescence, Zhou et al. find
that $T_{\rm BB}=0.4$~keV in the BB+PL fit, and $T_{\rm BB}=0.37$~keV
for the cooler component in the 2BB fit.  Within these observational
uncertainties, and accounting for the temperature corrections due to a
magnetized Hydrogen atmosphere, our theoretically predicted
temperature would be roughly consistent with the BB temperature
measured for this object in quiescence.
\begin{figure*}[t]
\begin{center}
\includegraphics[width=6.5in,angle=0]{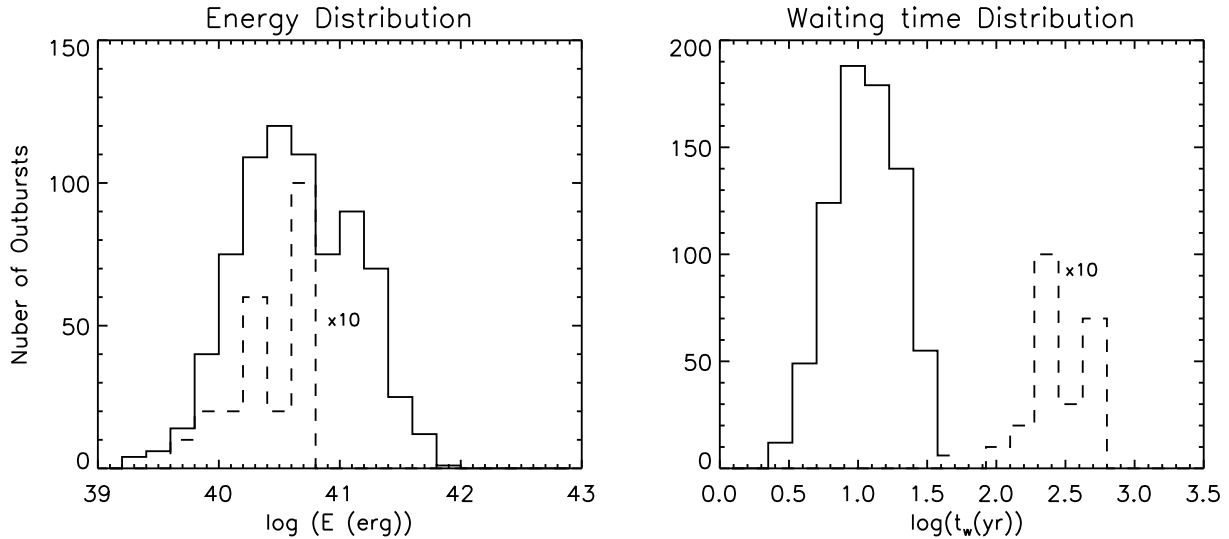}
\caption{Energy and waiting time distribution of the two cases representative
of \axp(solid) and \psr (dashes). The latter case is magnified a factor of 10 for comparison.
The total number of outbursts corresponds to a period of ten thousand years. 
\label{fig5}}
\end{center}
\end{figure*}
\noindent

Further constraints on our predictions for the thermal variables can
be made through a comparison with the pulse profile and pulsed
fraction of \axp. We used the predicted temperature distribution to
compute the expected flux from the pulsar during its rotation, and
hence calculate the pulse profile and the pulsed fractions in selected
energy bands.  For this calculation, we included the general
relativistic effects of gravitational redshift and light deflection,
following the formalism of Page (1995), with the modifications by
Perna \& Gotthelf (2008).  These allowed for the presence of a local
anisotropic radiation pattern, $f(\delta)\propto \cos\delta^n$
($\delta$ angle between the normal to the surface and the outgoing
photon direction, and $n$ a 'beaming' parameter, the only free
parameter that we used to find a match to the observed level of
modulation), as suggested by realistic atmospheric models.  Realistic
models were limited to fields perpendicular to the star surface,
(e.g. van Adelsberg \& Lai, 2006), and hence not directly applicable
to the spectrum from the entire NS surface. However, we still use them
to gauge the extent of the radiation beaming. Using the models of Van
Adelsberg \& Lai (2006), we find that, for $B\sim 10^{14}$~G and
$T\sim 0.4$~keV, the intensity for $E\sim 1$~keV drops as
$f(\delta)\propto \cos^2\delta$ up to $\delta\lesssim 50$~deg and
is shallower at larger $\delta$'s.  The effect of absorption to the
source (parameterized by the column density $N_{H}$) was also included
in our calculations, since it has been shown to be important for the
computation of the PF in finite energy bands (Perna et al. 2000).

We found the pulse profile to be double peaked, consistent with
observations (Woods et al. 2004).  We note that, for an orthogonal
rotator (and more generally for a large angle between the line of
sight and the rotation axis), the ``double peaked'' nature of the
pulse profile is a very robust prediction of our temperature profile,
since it is symmetric with respect to the equator. For n = 1.5, the
pulsed fractions produced by the temperature profile of our
simulations are 0.17 in the 0.3-1 keV band, and 0.22 in the 1-2 keV
band.These values comfortably sit within the measured quiescent values
for this source.  However, we need to note that, given the BB plus
powerlaw spectrum of this object, for the lowest energy band to be
dominated by the thermal component, the powerlaw needs to have a
cutoff in the soft X-ray band. This might be the case if hard photons
are due to upscattering of the thermal ones.  We did not attempt to
model the pulse profile in a quantitative fashion, given the
substantial contamination from the non-thermal component, for which no
predictions of its local angular distribution exist.  A full spectral
and timing analysis of the quiescent spectrum (as in Perna \& Gotthelf
2008 and Gotthelf et al. 2010) is beyond the scope of this paper, and
could only be performed with a better handle of the physical origin of
the powerlaw.

\bigskip
\noindent
{\em \underline{\psr}.}\\
\noindent

This object has a measured period $P=3.975$~s and period derivative
$\dot{P}=7.4\times 10^{-13}$~s/s (Camilo et al. 2000), yielding a
timing age of 85 kyr.  The dashed lines in Fig. \ref{fig4} show the
evolution of a similar object. The initial magnetic field set up in
this case is $B_{\rm p} = 1.6\times 10^{14}$~G and $B_{\rm tor}= 8\times
10^{14}$~G. At the measured spin-down age of 85~kyr, the real age of
the object is 60~kyr, with a predicted luminosity of about $4 \times
10^{33}$~erg/s, 5 times smaller than in the previous case. At the same
age, its period is 4.4~s, its spin-down rate $8 \times 10^{-13}$~s/s,
and its $B_{\rm p}$ about $9\times 10^{13}$~G.  This is again in good
qualitative agreement with the general properties of the object.  Note
that, for the X-ray luminosity, only upper limits exist.  An analysis
of {\em ROSAT} and {\em ASCA} archival data by Pivovaroff et
al. (2000) yielded the values $L_{0.1-2.4 {\rm keV}}< 6.3\times
10^{35}{d/10~{\rm kpc}}$~erg~s$^{-1}$ from {\em ROSAT} and $L_{2-10
  {\rm keV}}< 4.3\times 10^{33}{d/10~{\rm kpc}}$~erg~s$^{-1}$ from
the {\em ASCA} band.  Our predicted thermal (bolometric) luminosity is
consistent with these limits. Given the lack of X-ray detection, no 
comparison can be made at this stage with the pulse profile. 

With these results, we have answered one of the puzzles: despite the
second object under study has a similar dipolar field and is younger
than the first example, its luminosity is substantially lower. Indeed,
at other stages of evolution it can be even one order of magnitude
smaller.  The main reason for this difference, as we have discussed,
is the larger initial toroidal field of the first example.

Now we face the next problem: can the two objects have a markedly
different behaviour in their outbursting properties?  To answer this
question we have performed two simulations as described in Sect. 3,
following the evolution of the two objects for another $10^4$ yr but now
using the code including the Hall term in the induction equation. For
the same period of time, we found that the first object undergoes 753
starquakes, while the second one only had 23 of these events. On
average, one would expect that the first kind of source shows an
outburst every 10-20 years, while the second object only displays a
magnetar-like activity every 4-5 centuries. Our results are summarized
in Fig.~\ref{fig5}. In addition to the much lower event rate,
outbursts of objects like \psr\, are also less energetic than the
average magnetar outburst, thus making them harder to be observed.
Therefore, the apparent dichotomy of the two objects with similar
timing properties can easily be attributed to the slightly different
initial poloidal field and, more importantly, to the much stronger
internal toroidal field of one of the NSs.

\section{Summary}

In this work we have performed a comprehensive study of starquake
statistics on the NS crust, using the formalism developed by
\cite{pern11} for the coupled evolution of magnetic field and crust
stresses during the lifetime of the NS.  In particular, focusing on
objects at age $10^4-10^5$ yr, we have explored the dependence of the
outburst frequency and energetics on the initial strengths of the
dipolar and toroidal fields.  We have found that the outburst
frequency is negligible for initial poloidal fields below $10^{14}$~G,
even when the initial toroidal field is extremely strong, {$>
  10^{16}$~G }. The rate of outbursts is an increasing function of
$B_{\rm p}$ (at fixed $B_{\rm tor}$), and becomes significant (more than one
per century) for $B_{\rm p}(t=0)\ga 5\times 10^{14}$~G.  On the other hand,
at fixed $B_{\rm p}$, the initial toroidal field has practically no
influence on the starquake frequency for initial toroidal fields
$B_{\rm tor}\la B_{\rm p}$. This is because a toroidal field of the same
order of magnitude of $B_{\rm p}$ is anyway rapidly formed on a Hall
timescale.  For initial $B_{\rm tor}\ga B_{\rm p}$, the presence of the
toroidal component affects the outburst rate, producing its rapid
increase for initial $B_{\rm tor}$ much stronger than $B_{\rm p}$.

The relative strengths of $B_{\rm p}$ and $B_{\rm tor}$ at birth also
influence the thermal quiescent luminosity of the NS in a similar
fashion. For negligible $B_{\rm tor}$, the luminosity has a
strong dependence on $B_{\rm p}(t=0)$; it is about one order of magnitude
larger for $B_{\rm p}\sim 10^{15}$~G than it is for $B_{\rm p}\sim 10^{13}$~G. The
influence of the toroidal field becomes important, as for the outburst frequency, only for $B_{\rm
  tor}(t=0)\ga B_{\rm p}(t=0)$.

>From our simulations, we can conclude that there is a
critical value of the initial dipolar field strength, $B_{\rm p}(t=0)\sim 3-5 \times
10^{14}$~G, above which some regular active periods may be
expected. In this case, the intensity of the toroidal field at NS
birth plays a crucial role in the 'astrophysical appearance' of those
objects in their middle age.  In fact, at around $10^5$~yr, when $B_{\rm p}$
has roughly halved in strength to become $\sim 6-7\times 10^{13}$~G,
objects with negligible $B_{\rm tor}(t=0)$ have an outburst rate below
1 per century, making them appear more like a quiet ``high-$B$
pulsar''. On the other hand, with a larger toroidal field at birth, an
object with similar $B_{\rm p}$ can have an outburst rate of a few years,
making it appear as a 'magnetar'.  Correspondingly, due to the larger
toroidal field, the thermal luminosity of the 'magnetar' is higher
than that of the 'high-$B$ pulsar'. Interestingly, observations have
shown that middle-aged objects with measured $B_{\rm p}\sim 6-7\times
10^{13}$~G can display radically different thermal and bursting
properties. Specific examples are \axp\, and \psr: while they both
have $B_{\rm p}\sim 6\times 10^{13}$~G, the former manifests itself as a
magnetar, while the latter as a 'high-$B$ pulsar'. Here we have shown
how appropriate choices of $B_{\rm p}(t=0)$ and $B_{\rm tor}(t=0)$ can
naturally account for these differences, which no longer constitute a
puzzle within the NS phenomenology.

The apparent dichotomy between {\it quiet}, high B-field
radio-pulsars and {\it active} magnetars is not real. There is a continuum
of possibilities, and all types of sources can potentially show some
unusual activity (e.g. the recent case of SGR 0418+5729, Rea et al. 2010). 
We tend to classify them in classes according to a
biased version of the anthropic principle: if it has several active
periods on a human lifetime, we name it a magnetar; if it does not show an
outburst on the same timescale, we put it in the group of quiet, normal neutron stars. But
this separation is not necessarily associated to fundamental differences
in the nature of the sources or their internal physics.

\acknowledgements 
We thank an anonymous referee and Nanda Rea for helpful comments and suggestions.
This work was partly supported by CompStar, a Research Networking Programme of the European Science Foundation 
and grants  AYA2010-21097-C03-02, GVPROMETE02009-103 (JP) and 
NSF AST-1009396, NASA NNX10AK78G, NNX09AT17G, NNX09AT22G,
NNX09AU34G, GO0-11077X, DD1-12052X, G09-0156X, AR1-12003X, DD1-12053X (RP).%

\end{document}